\documentclass[aps,prl,twocolumn,superscriptaddress,showpacs]{revtex4}
\usepackage{graphicx}
\usepackage{dcolumn}
\usepackage{bm}

\usepackage{graphicx}
\usepackage{textcomp} 
\usepackage{amsmath}

\begin{document}

\title{Spatio-temporal coherent control of thermal excitations in solids}

\author{M. Sander}
\affiliation{Institute for Physics and Astronomy, Universit\"at Potsdam, Karl-Liebknecht-Str. 24-25, 14476 Potsdam, Germany}
\author{M. Herzog}
\affiliation{Institute for Physics and Astronomy, Universit\"at Potsdam, Karl-Liebknecht-Str. 24-25, 14476 Potsdam, Germany}
\author{J. E. Pudell}
\affiliation{Institute for Physics and Astronomy, Universit\"at Potsdam, Karl-Liebknecht-Str. 24-25, 14476 Potsdam, Germany}
\author{M. Bargheer}
\affiliation{Institute for Physics and Astronomy, Universit\"at Potsdam, Karl-Liebknecht-Str. 24-25, 14476 Potsdam, Germany}
\affiliation{Helmholtz-Zentrum Berlin for Materials and Energy GmbH, Wilhelm-Conrad-R\"ontgen Campus, BESSY II, Albert-Einstein-Str. 15, 12489 Berlin Germany}
\author{N. Weinkauf}
\affiliation{Institute for Solid State and Nanostructure Physics, Universit\"at Hamburg, Jungiusstr. 11, 20355 Hamburg, Germany}
\author{M. Pedersen}
\affiliation{European Synchrotron Radiation Facility ESRF, 71 Avenue des Martyrs 23800 Grenoble, France}
\author{G. Newby}
\affiliation{European Synchrotron Radiation Facility ESRF, 71 Avenue des Martyrs 23800 Grenoble, France}
\author{J. Sellmann}
\affiliation{Institute for Crystal Growth, Max-Born-Str. 2, 12489 Berlin, Germany}
\author{J. Schwarzkopf}
\affiliation{Institute for Crystal Growth, Max-Born-Str. 2, 12489 Berlin, Germany}
\author{V. Besse}
\affiliation{IMMM CNRS 6283, Universit\'e du Maine, 72085 Le Mans cedex, France}
\author{V. V. Temnov}
\affiliation{IMMM CNRS 6283, Universit\'e du Maine, 72085 Le Mans cedex, France}
\affiliation{Groupe d'Etude de la Mati\`ere Condens\'ee (GEMaC), Universit\'e de Versailles-Saint Quentin en Yvelines, CNRS UMR 8635, Universit\'e Paris-Sacley, 45 avenue des Etats-Unis, 78035 Versailles Cedex, France}
\author{P. Gaal}
\affiliation{Institute for Solid State and Nanostructure Physics, Universit\"at Hamburg, Jungiusstr. 11, 20355 Hamburg, Germany}
\affiliation{Helmholtz-Zentrum Berlin for Materials and Energy GmbH, Wilhelm-Conrad-R\"ontgen Campus, BESSY II, Albert-Einstein-Str. 15, 12489 Berlin Germany}

\date{\today}
\email{pgaal@physnet.uni-hamburg.de}

\begin{abstract}

X-ray reflectivity (XRR) measurements of femtosecond laser-induced transient gratings (TG) are applied to demonstrate the spatio-temporal coherent control of thermally induced surface deformations on ultrafast timescales. Using grazing incidence x-ray diffraction we unambiguously measure the amplitude of transient surface deformations with sub-\AA{} resolution. Understanding the dynamics of femtosecond TG excitations in terms of superposition of acoustic and thermal gratings makes it possible to develop new ways of coherent control in x-ray diffraction experiments. Being the dominant source of TG signal, the long-living thermal grating with spatial period $\Lambda$ can be canceled by a second, time-delayed TG excitation shifted by $\Lambda/2$. The ultimate speed limits of such an ultrafast x-ray shutter are inferred from the detailed analysis of thermal and acoustic dynamics in TG experiments.
\end{abstract}

\maketitle

Ultrafast photoacoustics\cite{Thomsen86PRB34,Ruel2015a} allows for the application of strain to a crystal lattice on ultrashort time- and lengthscales. Numerous studies have investigated coherent and incoherent phonon dynamics and the coupling of lattice deformations to electronic\cite{Weiß2014a}, optical\cite{Sing2014a}, magnetic \cite{Scherbakov2010,kim2012a,Thevenard10PRB82} or plasmonic \cite{TemnovNatureComm2013,TemnovJOPT2016} degrees of freedom. Strain-induced phenomena may be used to discover new material properties and develop new applications, for example the modification of optical and electronic properties in semiconductor nanostructures\cite{Wang2016a}. 
Surface acoustic waves (SAWs) are often employed as a source of lattice strain. They can be generated\cite{Roge2000a} and controlled\cite{Faye1986a} optically via the excitation of transient gratings (TGs)\cite{Vega2015a,Shen1997a}. Recently, these TG-excitations heave been used to probe heat transport in suspended thin films\cite{Vega2016a} and magneto-elastic coupling in thin nickel films \cite{Janu2016a,Janu2016b,ChangPRB2017}.
Optical excitation of a solid generates not only coherent sound waves but also incoherent thermal strain. Coherent excitations can be controlled in amplitude and phase by series of light pulses in time domain, which is labeled temporal coherent control\cite{Blan1997a}. The main fraction of the deposited optical energy is stored in incoherent excitations of the lattice, i.e., heat\cite{herz2012b, shay2011a} which can consequently not be controlled by a temporal sequence of light pulses. This thermal lattice excitation often generates a background which makes is difficult to precisely observe the coherent acoustic signal in purely optical experiments.

In this letter we demonstrate, for the first time, the coherent control of incoherent, thermal transient gratings. We apply spatio-temporal coherent control showing that the spatial part of coherent control adds a new degree of freedom to control the amplitude and the phase of a thermally deformed surface. This is clearly a new approach that introduces the concept of spatial coherent control to the dynamics of incoherent excitations on ultrafast time scales, a phenomenon impossible to achieve with temporal coherent control only. We also demonstrate the control of a transient thermal grating on a timescale faster than the oscillation of the simultaneously excited coherent acoustic modes. Our new quantitative method allows for decomposing the coherent and incoherent dynamics in the sample by measuring the amplitude of the surface excursion with sub-\AA~ precision and $\approx$~70 ps temporal resolution. The modification of x-ray diffraction intensity from laser-generated TGs is exploited to implement an ultrafast hard x-ray shutter. Whereas our data confirm the generation of Rayleigh SAWs and Surface Skimming Longitudinal Waves (SSLW)\cite{Janu2016b}, the most intriguing results deal with coherent control of the {\it incoherent, thermal} transient grating. The latter becomes possible by using a second time-delayed and spatially phase-shifted TG excitation.

\begin{figure}
  \centering
  \includegraphics[width = 0.5\textwidth]{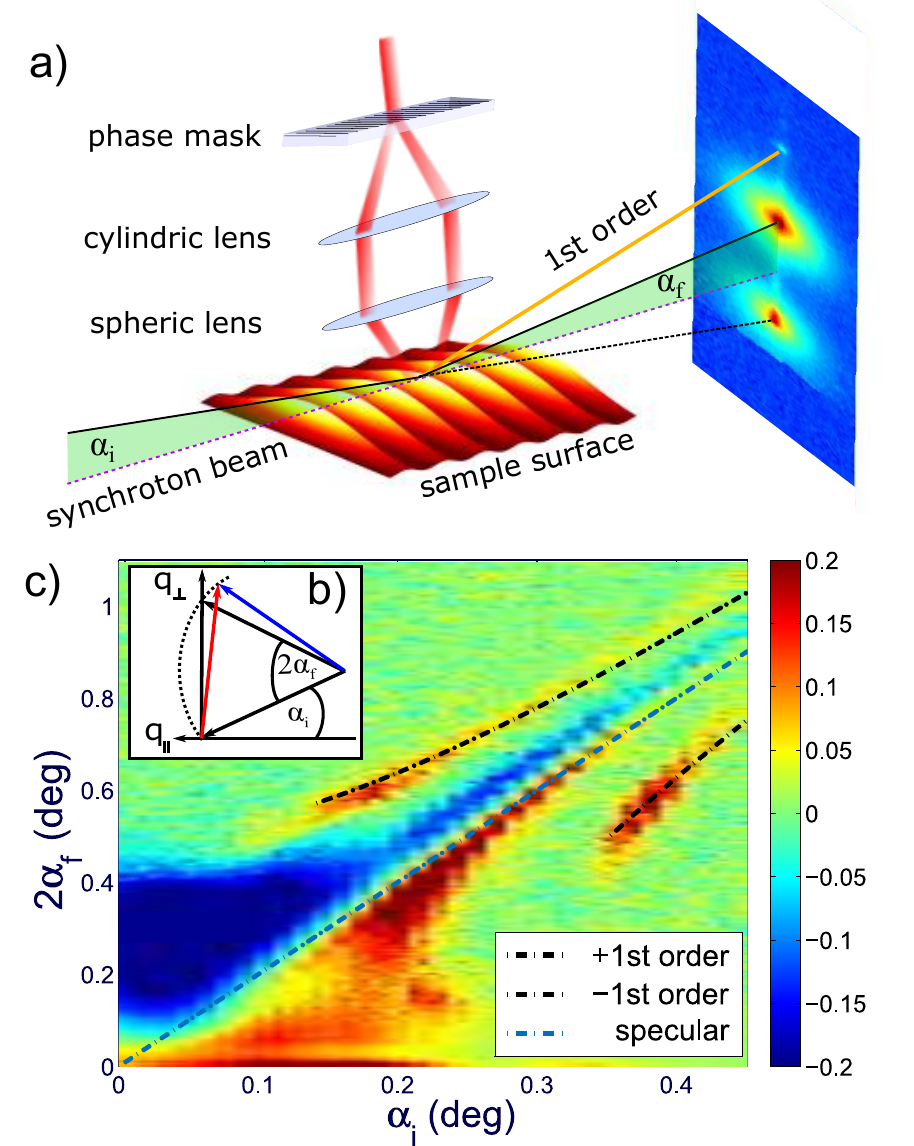}
  \caption{a) Experimental setup: 70\,ps x-ray pulses are delivered from the synchrotron storage ring of the European Synchrotron Radiation Facility ESRF. $\pm$1st diffraction orders of an ultrashort laser pulse are imaged onto the surface of the sample using a cylindrical and spherical lens in 4f-geometry and generate a spatially periodic distortion at the surface. The x-ray pulses impinge the sample at grazing incidence angles and diffract in a specular beam and $\pm$1st diffraction orders. A fraction of the direct beam is also visible at the bottom of the detector image. b) Ewald construction of the scattering vector (red arrow) c) Difference XRR measurement at a pump-probe delay $\tau$ of 500\,ps and a characteristic wavevector $\vec{q}_{||} = 2\pi/\Lambda$, where $\Lambda=8.8$~\textmu m. The measured specular reflection as well as $\pm$1st order correspond  to the theoretically expected positions (black and blue dashed lines).}
  \label{fig:ESRFsetup}
\end{figure}

\begin{figure}
  \centering
  \includegraphics[width = 0.4\textwidth]{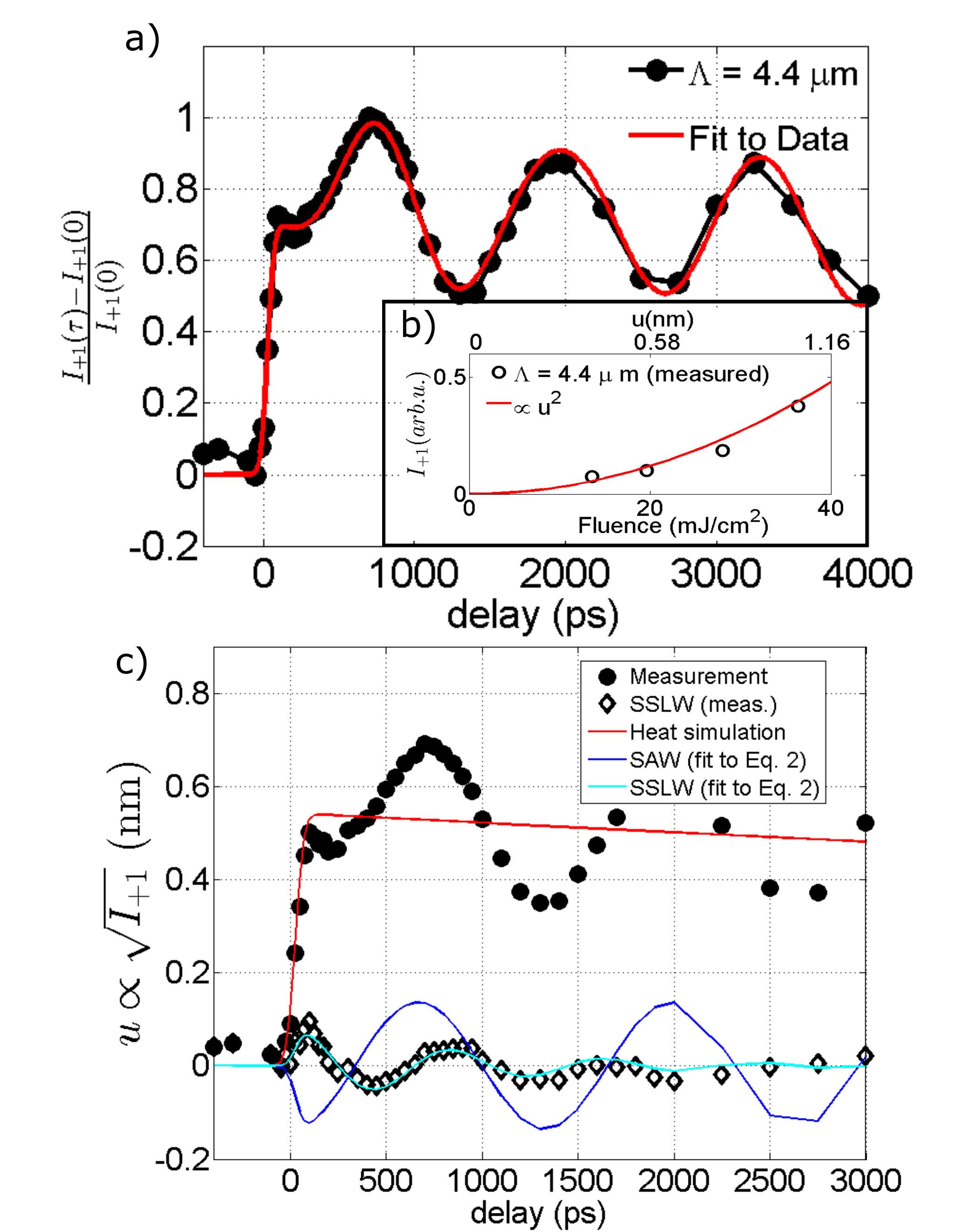}
  \caption{a) Time-resolved relative diffracted intensity in the +1st order $I_{+1}$ (black dots) and fit to the data using Eq.~\eqref{equ:SurfDistortion} (red solid line). The TG period was $\Lambda = 4.4$\,\textmu m. b) Calculated diffraction efficiency (symbols) vs.\ surface amplitude $u$. The diffraction model is described in the Methods section. The red solid line shows a quadratic surface amplitude dependence. c) Square root of the measured intensity $\sqrt{I_{+1}}$, which corresponds to the surface distortion (black dots). The y-scale is converted to a surface distortion amplitude with the empirical calibration factor $0.02 \frac{\%}{\text{mJ}/\text{cm}^{2}}$. The simulated thermal background, the SAW and the SSLW are shown in the red, blue and light blue solid lines, respectively. The black diamonds show the measured data after subtraction of thermal background and SAW.}
  \label{fig:Intensity}
\end{figure}

The optical TG excitation and experimental setup are shown in Fig.~\ref{fig:ESRFsetup}\,a). The sample was a 60\,nm Strontium Titanate (STO) on 150\,nm Strontium Ruthenate (SRO) thin film system grown on Dysprosium Scandate (DSO)\cite{Sell2014}.\@ The energy of the pump laser is absorbed only in the SRO film, which has a penetration depth of 44 nm, i.e., shorter than the film thickness. The photoacoustic properties of these materials are well understood. In particular, we studied nanoscale heat diffusion\cite{shay2011a}, thermoelastic effects\cite{Navirian2014} and coherent phonon dynamics\cite{herz2012c,Shayduk2013, Gaal2012, Gaal2014, sand2016a} in SRO, STO and similar materials. We assign a characteristic wavevector $|\vec{q}_{\parallel}| = 2\pi/\Lambda$ to the optically generated surface distortion with periodicity $\Lambda$. $\vec{q}_{||}$ is directed in the plane of the sample and in the diffraction plane of the incident x-ray beam. The periodic distortion results in a momentum transfer to the reflected x-ray beam $\vec{k}_{\text{out}}-\vec{k}_{\text{in}} = \vec{q}_{\perp}\pm\vec{q}_{\parallel}$. We emphasise the conceptional difference to inelastic x-ray diffraction by phonons\cite{boja2013a}: instead of a specific reciprocal lattice vector $\vec{G}$, $\vec{q}_{\perp}$ denotes the recoil momentum due to specular reflection. It is directed perpendicular to the sample surface and of continuous magnitude $\vec{q}_{\perp} = 2k_{\text{in}}\sin{\alpha_{\text{i}}}$. The Ewald construction of the scattering vector is shown in Fig.~\ref{fig:ESRFsetup}\,b). It reveals that the x-ray photons diffracted by the periodic surface excursion (PSE) exhibit an exit angle $\alpha_{\text f} \neq \alpha_{\text i}$. Indeed, a typical detector image presented in Fig.~\ref{fig:ESRFsetup}\,a) shows a first-order diffraction spot (orange solid line) above the specular total reflection (black solid line). Fig.~\ref{fig:ESRFsetup}\,c) shows a differential x-ray reflectivity measurement $R(\tau, \alpha_{\text i})-R(-\infty,\alpha_{\text i})$ at a pump-probe delay of $\tau$ = 500\,ps around the critical angle $\alpha_{\text c}$ of total reflection. Evaluation of the scattering condition for a spatial period of $\Lambda = 8.8$\,\textmu m yields the blue and black dashed curves which are in excellent agreement with the angular position of the specular and $\pm$1st-order reflection, respectively. Having established the capability of picosecond x-ray reflectivity measurements to detect a PSE and to quantify its spatial periodicity, we now investigate the dynamics of impulsively excited PSEs. The black symbols in Fig.~\ref{fig:Intensity}\,a) represent the time-dependent intensity of the first-order diffraction peak $I_{+1}(\tau)$ at $\alpha_{\text i}=0.16^{\circ}$ after TG excitation with a period of $\Lambda=4.4$~\textmu m and an incident fluence of 19.6\,mJ/cm$^{2}$. Note that $\alpha_{\text{i}}<\alpha_{\text{c}}$, i.e., only the evanescent field of the incident x-ray beam penetrates the sample and the main part of the beam is diffracted from the surface\cite{warren1969}.  Upon TG excitation ($\tau=0$) we observe a step-like increase of the diffracted intensity. The initial step, which occurs with the temporal resolution of the experiment, is followed by a signal decrease that lasts for 300\,ps. Subsequently we observe oscillations of the diffracted intensity with a period of $T_{\text{SAW}} = 1310$\,ps.
 Due to the short x-ray penetration depth in our surface diffraction experiment, we rely on kinematic theory \cite{Madsen2005a, Nicolas2014a, Irzhak2014, Roshchupkin2009} and find the following expression for the diffracted intensity from a PSE under grazing incidence below the critical angle $\alpha_i < \alpha_c$:
\begin{eqnarray} \label{equ:DiffIntegral}
 I_{1} & = &\left|\int_{\parallel} e^{-i\left(q_{\parallel}\cdot r_{\parallel}+\frac{\Delta\varphi}{2}\sin(\frac{2\pi}{\Lambda}r_{\parallel})\right)}dr_{\parallel}\right|^{2} \\ \nonumber
  & =  & \left|J_{1}\left(\frac{\Delta\varphi}{2}\right)\right|^{2}
\end{eqnarray}
$J_{1}(\frac{\Delta\varphi}{2})$ is the first order Bessel function and ${\Delta\varphi}$ denotes the phase difference between x-ray wave fields reflected from the maximum and the minimum of the PSE. For small arguments $\frac{\Delta\varphi}{2}$, $I_{1}$ is proportional to the square of the time-dependent surface amplitude $u(\tau)$, as evidenced by Fig.~\ref{fig:Intensity}\,b). Hence, in the limit of small deformations, the x-ray intensity diffracted from the PSE is directly proportional to the squared surface amplitude $I_{+1}\propto u(\tau)^{2}$. A quantitative comparison of experimental data with theoretical models\cite{Herzog2012, herz2012b, Schick2014a} shows that an optical excitation fluence of 1\,mJ/cm$^{2}$ results in a lattice strain of 0.2$\pm$0.05\text{\textperthousand}. We derive the calibration factor in the Supplemental Material\cite{Supplement}. With our excitation fluence of 19.6\,mJ/cm$^{2}$ and a thickness of 150\,nm of the absorbing SRO layer, we determine the maximum surface amplitude to 0.57\,nm. The y-scale of Fig.~\ref{fig:Intensity}\,c) is an absolute scale which quantifies the surface distortion in nanometers.

\begin{figure}
  \centering
  \includegraphics[width = 0.45\textwidth]{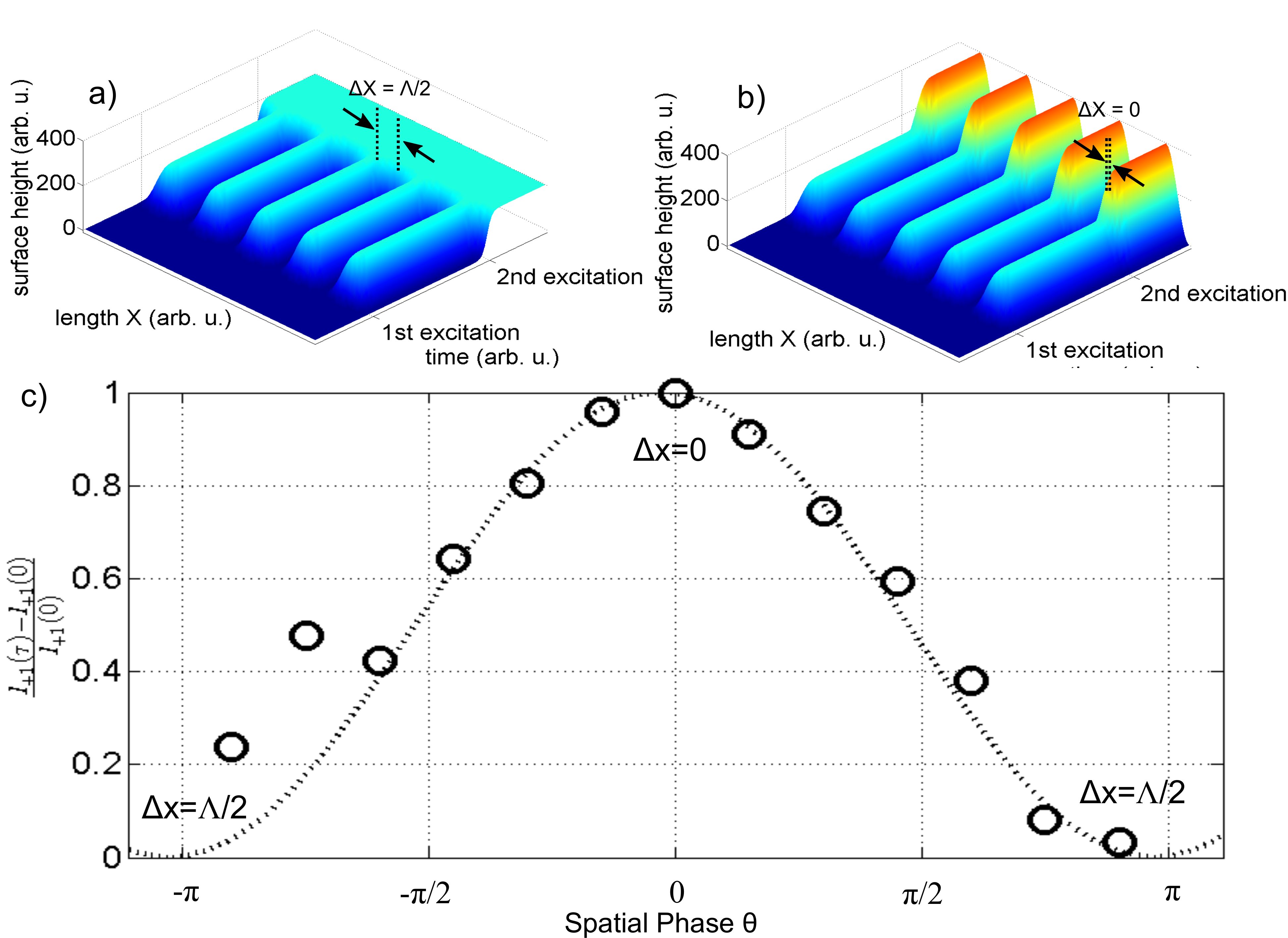}
  \caption{Coherent control of a {\it thermal} transient grating using two TG excitations. a) and b) The first excitation generates a thermal grating at $\tau = 0$. The second excitation at $\tau_{12} = 200$\,ps creates another thermal grating shifted by $\Delta x=\Lambda/2$ (a) and $\Delta x = 0$ (b), thus canceling (enhancing) the first grating. c) Diffracted intensity I$_{+1}$ at a pump-probe delay of $\tau = \tau_{12} + 150$\,ps vs.\ the spatial phase $\Theta$. The second TG excitation impinges the sample at a delay of $\tau_{12} = 200$\,ps.}
  \label{fig:2Pexp}
\end{figure}

Impulsive optical excitation of an absorbing medium with a spatially periodic intensity distribution results in a time-dependent surface deformation $u(\tau)$ of the form: $ u(\tau) = u_{\text{h}}\exp(-\alpha_{\text{h}}\tau) + u_{\text{c}}(\tau)$,
where the the first term accounts for the thermal grating. The coherent part $u_{\text{c}}$ is the sum of SAW and SSLW contributions, respectively \cite{Janu2016b}:
\begin{eqnarray} \label{equ:SurfDistortion}
  u_{\rm c}(\tau) & = & u_{\rm SAW}\cdot\cos(\omega_{\rm SAW}\cdot\tau+\varphi_{\rm SAW}) \\ \nonumber
  & + & u_{\rm SSLW}\cdot\cos(\omega_{\rm SSLW}\cdot\tau + \varphi_{\rm SSLW})\cdot {\rm exp}(-\alpha_{\rm SSLW}\tau)\,
\end{eqnarray}
With the diffraction model outlined above, we can interpret all components that result from the decomposition of the measured surface amplitude $u(\tau)$ shown in the black dotted curve in Fig.~\ref{fig:Intensity}\,c). To model the thermal background $u_{\text{h}}(\tau)$ of the surface amplitude, we numerically solve the two-dimensional heat diffusion equation for one period of the PSE with periodic boundary conditions. The initial temperature profile into the depth of the absorbing SRO thin film is determined by the optical penetration depth and decays exponentially. We choose the in-plane and out-of-plane thermal conductance of SRO and DSO in order to reproduce the slow decay of the incoherent signal contribution (see Supplemental Material\cite{Supplement}).
By subtracting the simulated thermal background [red solid line in Fig.~\ref{fig:Intensity}\,c)] we obtain the coherent part of the time-dependent surface amplitude, which we decompose in two modes by fitting Eq.~\eqref{equ:SurfDistortion} to our data. The blue solid line in Fig.~\ref{fig:Intensity}\,c) represents the amplitude $u_{SAW}$ of the Rayleigh-wave propagating parallel the surface. Further subtraction of the Rayleigh wave from the coherent signal reveals a strongly damped mode with a decay constant of 500\,ps and a period of $(T_{\text{SSLW}}=800$\,ps shown by the black diamonds (measurement) and the light blue solid curve (fit to data) in Fig.~\ref{fig:Intensity}\,c). We assign this feature to the Surface Skimming Longitudinal Wave (SSLW) that is generated by the optical excitation and propagates into the substrate.
The results in Fig. 2 clearly demonstrate the ability to decompose the full surface dynamics into individual coherent acoustic and incoherent thermal components. In the following we adress the incoherent thermal background to fully control the surface deformation in a coherent control scheme. In particular our experiment entails controlling the lateral heat flow in the sample by a series of consecutive TG excitations with defined time delay $\tau_{12}$. In addition, we control the spatial phase $\Theta=\frac{2\pi\Delta x}{\Lambda}$ by adjusting the angle of incidence of the second pulse on the transmission phase mask (see Methods section). $\Delta x$ denotes the distance between the maxima of first and second TG excitation. Examples for $\Theta = \pi$ and $\Theta = 0$ are shown in Fig.~\ref{fig:2Pexp}\,a) and b) where the PSE is eliminated and enhanced , respectively.
An experimental demonstration of this novel spatio-temporal TG coherent control is shown in Fig.~\ref{fig:2Pexp}\,c), where the first-order diffracted intensity at a delay $\tau = \tau_{12} + 150$\,ps  is plotted as a function of spatial phase $\Theta$. The delay between the first and second TG excitation $\tau_{12}$ is 200\,ps. The experimental data follow the expected $(1+\cos\Theta)$-dependence (dashed line) and thus evidence spatial coherent control of a thermal grating on ultrafast timescales.

Finally we demonstrate coherent control of the thermal grating along the temporal degree of freedom by keeping the spatial phase $\Theta=\pm \pi$ constant and by changing the time delay $\tau_{12}$ between the first and second TG excitation. We restrict the experiment to pump-probe delays much shorter than the period of the coherent modes, i.e., $\tau_{12} \ll T_{SAW}, T_{SSLW}$. Hence the temporal phase difference of the coherent modes launched by either TG excitation nearly vanishes, because the spatial alignment introduces a phase shift of $\pi$. The total spatio-temporal phase difference results in complete destructive interference of {\it both} the thermal grating and of the coherent sound waves.
\begin{figure}
  \centering
  \includegraphics[width = 0.5\textwidth]{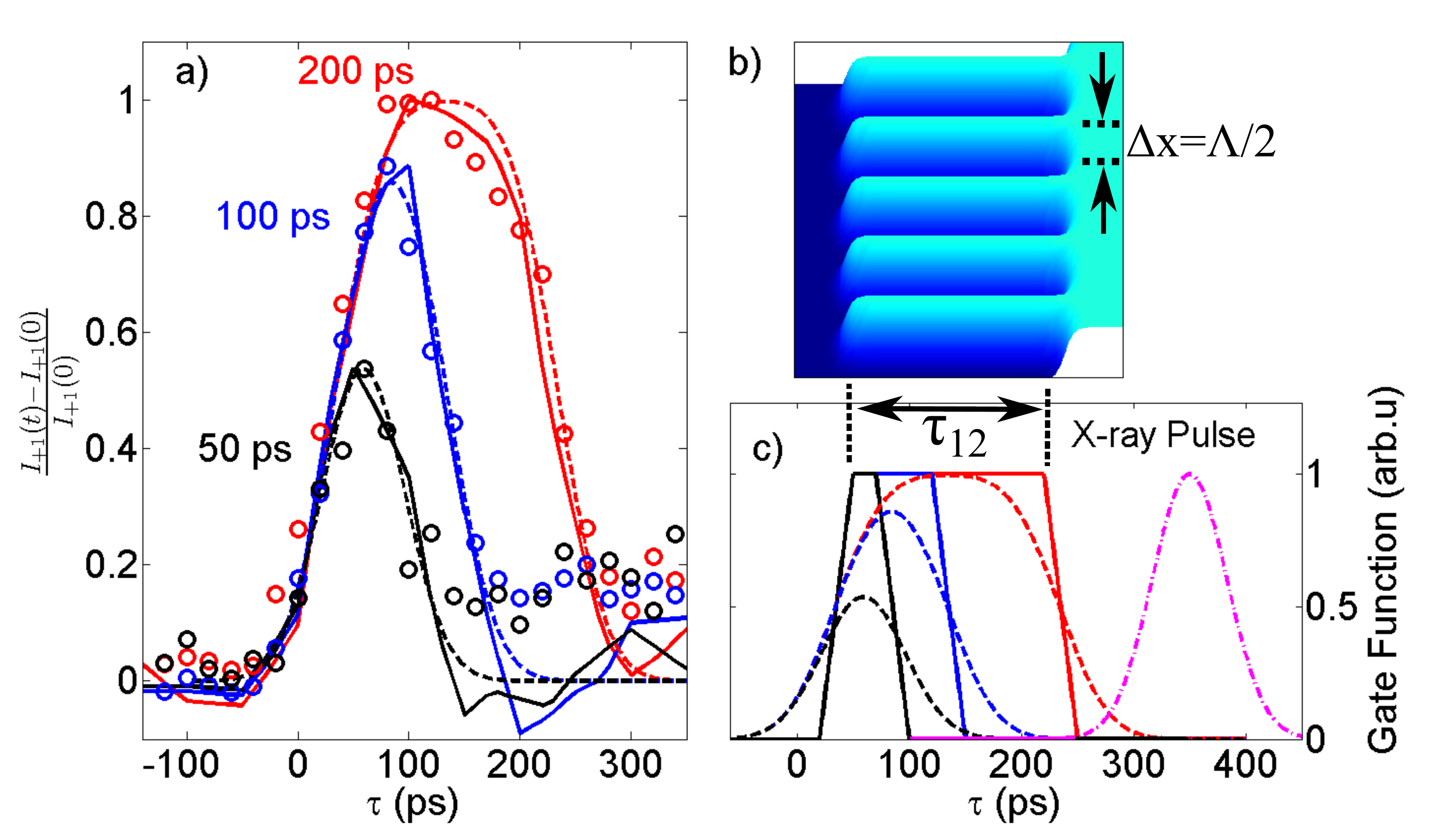}
  \caption{Spatio-temporal control of the thermal and coherent surface grating. The relative delay between first and second TG excitation $\tau_{12}$ is 200\,ps (red), 100\,ps (blue) and 50\,ps (black), respectively. a) Schematic view of the transient surface distortion. b) Construction of the gate function (dashed lines) by convolution of the incident x-ray pulse (pink dash-dotted line) with the gate that is spanned by two TG excitations. c) Comparison of measurement (symbols) and theoretical gate functions (solid and dashed lines).}
  \label{fig:2Pexp_TD}
\end{figure}

The above considerations are fully confirmed by the experimental data (symbols) shown in Fig.~\ref{fig:2Pexp_TD}\,a) which evidence an almost complete elimination of the first-order diffraction efficiency by the second TG excitation for time delays $\tau_{12}=50, 100, 200$\,ps. In particular, all traces reveal that the transient PSE opens a temporal gate for incident x-ray photons that is spanned by the first and second TG excitation. The duration of the gate can be controlled by $\tau_{12}$ even to durations shorter than the incident x-ray pulse. However, the measured gate width cannot be shorter than the duration of the x-ray probe pulse \cite{Gaal2012,sand2016a} although individual x-ray pulses are truncated by the gate. This reduces the signal strength and decreases the signal-to-noise ratio in this measurement.
To prove that the recorded data shown in Fig.~\ref{fig:2Pexp_TD}\,a) evidence the coherent superposition of two time-delayed laser-induced surface gratings, we construct the corresponding gate function by summation of two replica of the measured transients for a single-TG excitation such as shown in Fig.~\ref{fig:Intensity} with appropriate relative sign and time delay. For a spatial phase $\Theta=\pm\pi$ the second TG excitation reduces the amplitude of the surface deformation, which we incorporate by a negative sign of the second surface amplitude response. We further add the relative time delay $\tau_{12}$ to the second transient. This approach, which is explained in more detail in the Supplemental Material\cite{Supplement}, is solely based on experimentally measured transients and yields the solid lines in Fig.~\ref{fig:2Pexp_TD}\,a). We observe a very good agreement between the measured and the constructed gate functions. The dashed lines in Fig.~\ref{fig:2Pexp_TD}\,a) result from a convolution of gate functions shown in Fig.~\ref{fig:2Pexp_TD}\,c) with a Gaussian-shaped x-ray probe pulse. This simple model also yields very good representation of the experimental data.

In conclusion, we demonstrated full control over a thermally excited incoherent surface deformation by introducing spatio-temporal coherent control in ultrafast laser-induced TG experiments. Our new approach allows for a separation of the coherent and incoherent response of the optically generated lattice excitation. We show that the incoherent thermal grating can be controlled on a timescale significantly shorter than the oscillation period of the coherent modes, which we exploit to implement an ultrafast shutter in gracing incidence x-ray diffraction geometry. The presented concept is not limited to thin metallic transducers deposited on dielectric substrates: it will work on any optically opaque bulk sample. However, the ultimate speed of an x-ray shutter will be determined by the acoustic propagation time through the depth of the laser-generated stress. With the demonstrated outstanding sub-angstrom spatial precision and a sub-100~ps instrumental time resolution we are looking forward towards applying an ultrafast x-ray gating technique to resolve the dynamics of strain-and heat-induced phenomena in solids and nanostructures.

The authors would like to acknowledge valuable discussion with Flavio Zamponi, support during the beamtime by Michael Wulff and financial support from BMBF via 05K16GU3 and from {\it Strat\'{e}gie Internationale} "NNN-Telecom" de la R\'{e}gion Pays de La Loire, ANR-DFG "PPMI-NANO" (ANR-15-CE24-0032 \& DFG SE2443/2).


\end{document}